# Minimal weak basis textures and quark mixing data

Rohit Verma

*Rayat Institute of Engineering and Information Technology, Ropar, India*

*Email: rohitverma@live.com*

Sept 26, 2013

**Abstract**

The phenomenology of most general quark mass matrices with minimum of three texture zeros obtained through weak basis (WB) transformations has been discussed within the framework of the Standard Model (SM). The relationship of textures, naturalness and weak basis transformations has also been studied. Adopting a 'hierarchical' parameterization for the quark mass matrix, we have detailed exact relations for the various elements of the quark mixing matrix along with the CP asymmetry angle β illustrating the explicit relationship between the observed hierarchies among the WB quark mass matrices and the quark mixing matrix. Besides exploring the complete parameter space available to the $(M_q)_{11}$, (q = u, d) elements in these mass matrices, their implications for quark mixing data and CP violation asymmetries have also been explored.

## 1. Introduction

In the SM framework, the quark mass matrices are arbitrary 3 × 3 complex matrices involving 36 free parameters. In the absence of flavor-changing right-handed currents, one can always redefine the right-handed quark fields such that the resulting quark mass matrices are hermitian in general. However, one still has the freedom to apply an overall unitary transformation on the left-handed fields which does not change the charged currents and therefore are not measureable with the Standard Model interactions. Such unitary transformations are the so-called weak basis transformations and do not affect the hermicity of the mass matrices and thus the number of free parameters in these is reduced to 18. This number is still very large as compared to only 10 physical observables corresponding to 6 quark masses and 4 physical parameters of the CKM matrix [1] viz. the three mixing angles and one CP violation phase.

In view of this, several phenomenological "Bottom- Up" approaches have been widely adopted so as to extract information regarding the structural features of these matrices using the experimental low energy data. Broadly, these include [2] radiative mechanisms, texture zeros, family symmetries and seesaw mechanisms. One of the successful ansätze incorporating the 'texture zero' approach was initiated by Weinberg and Fritzsch [3].

*A particular texture structure is said to be texture n zero, if it has n number of non-trivial zeros, for example, if the sum of the number of diagonal zeros and half the number of the symmetrically placed off diagonal zeros is n.* The texture based approach facilitates in relating the elements of the quark mixing (CKM) matrix with those of the mass matrices as well as with the quark masses.

It was observed [4], [5] that using the freedom of weak basis (WB) transformations, it is possible to obtain hermitian quark mass matrices involving a 'maximum' of 3 phenomenological texture zeros 'without' having any 'physical' implications. Such a unitary transformation preserves the hermicity of the mass matrices as well as leaves the weak currents invariant. Any 'additional' texture zero is supposed to have physical implications. But the maximum allowed number of such phenomenological texture zeros is just six, with three in the up and three in the down quark sectors respectively, corresponding to diagonal quark mass matrices.

Among the several possibilities of texture specific mass matrices, all the texture 6 zero quark mass matrices are known to be ruled out entirely by the current experimental data [6]. However, although some of the texture 5 zero quark mass matrices are able to explain the mixing data [7], though not completely explaining the CP violation phenomenon at 1σ C.L., the Fritzsch-like texture 4 zero hermitian mass matrices appear to be most promising and have been widely discussed in the literature [4],[5],[8],[9]. Since the latter involve 4 texture zeros, these are not usually considered as the most general quark mass matrices in view of the weak basis transformations.

From our current understanding of the quark mixing phenomenon and its relation to phenomenological texture specific hermitian quark mass matrices, it is widely accepted that the diagonal elements in these mass matrices do not exhibit a strong hierarchy i.e. $(M_q)_{22} \gg m_2$ [4]. However, a thorough investigation of the comprehensive mathematical formulae signifying the exact relationship of observed hierarchies among the quark mass matrices and the CKM matrix elements is still unexplored in the literature. One is usually encountered with complicated mathematical relations for the various CKM elements which are difficult to comprehend [2], [4]. The purpose of the current manuscript on one hand is to derive the precise relations for the various CKM elements wherein the effect of hierarchy in the mass matrices on the mixing phenomenon is clearly evident, while on the other hand, we investigate the implications of the *additional* non-zero elements in the texture 3 zero hermitian quark mass matrices, emerging from WB transformations, for the quark flavor mixing data and CP violation asymmetries.

## 2. WB Transformations and Minimal textures

Starting with the basis wherein one of the quark mass matrices among $M_u$ and $M_d$ is real diagonal while the other is an arbitrary hermitian matrix, in the *u-diagonal representation* [7] where $M_u$ is real diagonal, we have

$$M_u = \begin{pmatrix} m_u & 0 & 0 \\ 0 & m_c & 0 \\ 0 & 0 & m_t \end{pmatrix}, \quad M_d = V \begin{pmatrix} m_d & 0 & 0 \\ 0 & m_s & 0 \\ 0 & 0 & m_b \end{pmatrix} V^\dagger = \begin{pmatrix} e_d & |a_d|e^{i\alpha_d} & |f_d|e^{i\omega_d} \\ |a_d|e^{-i\alpha_d} & d_d & |b_d|e^{i\beta_d} \\ |f_d|e^{-i\omega_d} & |b_d|e^{-i\beta_d} & c_d \end{pmatrix}, \quad (1)$$

and V is the unitary CKM matrix. In order to obtain clues for the structure of the matrix $M_d$ in this representation, we consider V to be real for simplicity. It is clear from Eqn. (1) that the matrix $M_u$ has three texture zeros and $M_d$ has no texture zero; such a combination of $M_u$ and $M_d$ is usually referred to as a texture 3 zero case. In order to obtain the general structure of the mass matrix $M_d$ in the *u-diagonal representation*, we make use of Eqn. (1) and take $m_d$ = 2.9 MeV, $m_s$ = 55 MeV, $m_b$ = 2.9 GeV from [10] and V from [11] viz.

$$V = \begin{pmatrix} 0.97427 \pm 0.00015 & 0.22534 \pm 0.00065 & 0.00351^{+0.00015}_{-0.00014} \\ 0.22520 \pm 0.00065 & 0.97344 \pm 0.00016 & 0.0412^{+0.0011}_{-0.0005} \\ 0.00867^{+0.00029}_{-0.00031} & 0.0404^{+0.0011}_{-0.0005} & 0.999146^{+0.000021}_{-0.000046} \end{pmatrix}, \quad (2)$$

and obtain, for example

$$|M_d| = \begin{pmatrix} 0.0055 & 0.0131 & 0.0107 \\ 0.0131 & 0.0572 & 0.1215 \\ 0.0107 & 0.1215 & 2.8948 \end{pmatrix} \text{GeV}. \quad (3)$$

In case one gives a complete variation to the quark masses and CKM elements above, for both the *u-diagonal* and *d-diagonal* representations, one obtains

$$e_q < |a_q|, |f_q| < d_q < |b_q| < c_q. \quad (4)$$

It is interesting to note that for the quark sector, the observed hierarchy among the quark masses i.e. $m_1 \ll m_2 \ll m_3$ as well as CKM elements i.e. $(V_{ub}, V_{td}) < (V_{cb}, V_{ts}) < (V_{us}, V_{cd}) < (V_{ud}, V_{cs}, V_{tb})$ gets naturally translated on the structure of the corresponding quark mass matrices. Such hierarchical mass matrices have been referred to in the literature as *natural mass matrices* [12]. However, in principle an exact analytical diagonalization of the matrix $M_d$ given in Eqn. (1) is not always possible, making it difficult to correlate the quark mixing parameters with the elements of the mass matrices.

In view of this, one can apply a WB transformation, which is essentially a unitary transformation acting simultaneously on the mass matrices $M_u$ and $M_d$ defined in Eqn. (1) such that

$$M_u \to M'_u = U M_u U^\dagger, \qquad M_d \to M'_d = U M_d U^\dagger, \tag{5}$$

where U is an arbitrary unitary matrix. It is easy to check that the two representations $(M_u, M_d)$ and $(M'_u, M'_d)$ are physically equivalent in the sense that they lead to the same CKM matrix. The complete proof of the existence of such bases was first provided quantitatively by Branco *et al.* [5] wherein both diagonal and non-diagonal texture zeroes were obtained using the freedom of WB transformations. However, Fritzsch *et al.* [4] have qualitatively justified the non-diagonal texture zeroes originating from such transformations. As discussed in reference [4] there is a possible choice of U such that

$$(M'_u)_{13,31} = (M'_d)_{13,31} = (M'_d)_{11} = 0 \tag{6}$$

or

$$(M'_d)_{13,31} = (M'_u)_{13,31} = (M'_u)_{11} = 0 \tag{7}$$

and other elements of these matrices non-zero. This corresponds to either $M'_u$ being a WB texture 1 zero hermitian mass matrix and $M'_d$ being a WB texture 2 zero hermitian mass matrix in Eqn. (6) or vice versa in Eqn. (7). It is interesting to note that although the total number of texture zeros are equal to three in both the representations $(M_u, M_d)$ and $(M'_u, M'_d)$, the texture structures of the corresponding mass matrices are different. It so appears that the application of the WB transformation leads to a redistribution of texture zeros in the two mass matrices, thus altering their structural form. It is therefore desirable to check if the mass matrices in the representation $(M'_u, M'_d)$ also follow the condition of naturalness given by Eqn. (4).

In the contrary, Fritzsch-like texture 4 zero hermitian mass matrices are characterized by

$$(M'_u)_{13,31} = (M'_u)_{11} = (M'_d)_{13,31} = (M'_d)_{11} = 0, \tag{8}$$

with other elements being non-zero. The WB texture 3 zero hermitian mass matrices in Eqns. (6) and (7) differ from the Fritzsch-like texture 4 zero hermitian mass matrices in Eqn. (8) in the sense that $(M'_u)_{11} \neq 0$ in Eqn. (6) and $(M'_d)_{11} \neq 0$ in Eqn. (7). This necessitates an investigation of the implications of the *additional* diagonal matrix elements $(M'_u)_{11} = e_u$ and $(M'_d)_{11} = e_d$ in the matrices $(M'_u, M'_d)$, obtained through a WB transformation, for quark mixing and CP violation phenomenon, vis-à-vis Fritzsch-like texture 4 zero hermitian quark mass matrices. Recently, it was observed [4] that the latter exhibit naturalness of Eqn. (4). It is therefore desirable to check the complete parameter space available to the elements $e_u$ and $e_d$ in WB texture 3 zero mass matrices $(M'_u, M'_d)$ as well as the extent of dependence of the quark mixing parameters and CP violation asymmetries on these.

## 3. Diagonalization of WB textures

Note that for the WB textures of quark mass matrices mentioned in Eqns. (6) and (7), one of the mass matrix is a Fritzsch-like texture 2 zero hermitian mass matrix whereas the other has the following form

$$M'_q = \begin{pmatrix} e_q & |a_q|e^{i\alpha_q} & 0 \\ |a_q|e^{-i\alpha_q} & d_q & |b_q|e^{i\beta_q} \\ 0 & |b_q|e^{-i\beta_q} & c_q \end{pmatrix}, \quad q = u, d. \tag{9}$$

It may be noted that in comparison with Fritzsch-like texture 2 zero mass matrix given in Eqn. (8), the above mass matrix has an additional non-zero diagonal element $(M'_q)_{11} = e_q$. To find the allowed parameter space of this parameter as well as to construct the mixing matrix, one needs to obtain its diagonalizing transformation first. It is possible to arrive at the following equations relating the mass matrix elements $c_q$, $a_q$ and $b_q$ with the quark masses $m_1$, $m_2$, $m_3$ and the free parameters $d_q$ and $e_q$ [13]

$$c_q = m_1 - m_2 + m_3 - d_q - e_q, \quad |a_q| = \sqrt{\frac{(m_1 - e_q)(m_2 + e_q)(m_3 - e_q)}{(c_q - e_q)}},$$

$$|b_q| = \sqrt{\frac{(c_q - m_1)(m_3 - c_q)(c_q + m_2)}{(c_q - e_q)}}. \tag{10}$$

In order that $|a_q|$ and $|b_q|$ remain real, the free parameters $e_q$ and $d_q$ get constrained within the limits

$$m_1 > e_q > -m_2, \quad (m_3 - m_2 - e_q) > d_q > (m_1 - m_2 - e_q). \tag{11}$$

This indicates that the condition of hermicity on the texture 1 zero mass matrix in Eqn. (9) restricts the parameter $e_q$ to have very small values only, thereby satisfying the naturalness condition discussed in Eqn. (4). The exact diagonalizing orthogonal matrix for the real mass matrix $M^r_q$, defined as $M'_q = P^\dagger_q M^r_q P_q$ where $P_q = \text{Diag}\{e^{-i\alpha_q}, 1, e^{i\beta_q}\}$, is given below

$$O_q = \begin{pmatrix} \sqrt{\dfrac{(e_q+m_2)(m_3-e_q)(c_q-m_1)}{(c_q-e_q)(m_3-m_1)(m_2+m_1)}} & \sqrt{\dfrac{(m_1-e_q)(m_3-e_q)(c_q+m_2)}{(c_q-e_q)(m_3+m_2)(m_2+m_1)}} & \sqrt{\dfrac{(m_1-e_q)(e_q+m_2)(m_3-c_q)}{(c_q-e_q)(m_3+m_2)(m_3-m_1)}} \\ \sqrt{\dfrac{(m_1-e_q)(c_q-m_1)}{(m_3-m_1)(m_2+m_1)}} & -\sqrt{\dfrac{(e_q+m_2)(c_q+m_2)}{(m_3+m_2)(m_2+m_1)}} & \sqrt{\dfrac{(m_3-e_q)(m_3-c_q)}{(m_3+m_2)(m_3-m_1)}} \\ -\sqrt{\dfrac{(m_1-e_q)(m_3-c_q)(c_q+m_2)}{(c_q-e_q)(m_3-m_1)(m_2+m_1)}} & \sqrt{\dfrac{(e_q+m_2)(c_q-m_1)(m_3-c_q)}{(c_q-e_q)(m_3+m_2)(m_2+m_1)}} & \sqrt{\dfrac{(m_3-e_q)(c_q-m_1)(c_q+m_2)}{(c_q-e_q)(m_3+m_2)(m_3-m_1)}} \end{pmatrix}. \tag{12}$$

In order to simplify the above matrix, one can make use of the strong hierarchy among the quark masses i.e. $m_1 \ll m_2 \ll m_3$ and obtain

$$O_q = \begin{pmatrix} 1 & \sigma_q\sqrt{\dfrac{m_1}{m_2}}\sqrt{1-\xi_q} & \dfrac{m_1 m_2}{m_3^2}\sqrt{(\zeta_q + m_2/m_3)}\sqrt{1-\xi_q} \\ \sqrt{\dfrac{m_1}{m_2}}\sqrt{\dfrac{1-\xi_q}{1+\zeta_q}} & -\sigma_q\sqrt{\dfrac{1}{1+\zeta_q}} & \sqrt{\dfrac{(\zeta_q + m_2/m_3)}{1+\zeta_q}} \\ -\sigma_q\sqrt{\dfrac{m_1}{m_2}}\sqrt{\dfrac{(\zeta_q + m_2/m_3)}{1+\zeta_q}}\sqrt{1-\xi_q} & \sqrt{\dfrac{(\zeta_q + m_2/m_3)}{1+\zeta_q}} & \sigma_q\sqrt{\dfrac{1}{1+\zeta_q}} \end{pmatrix}, \quad (13)$$

where the free parameters $\xi_q$ and $\zeta_q$ are the hierarchy characterizing parameters for the mass matrix $M_q$ and are defined as

$$\xi_q = e_q/m_1 \text{ and } \zeta_q = d_q/c_q, \quad (14)$$

and

$$\sigma_q = \sqrt{1+\left(\dfrac{m_2}{m_3}(1+\zeta_q)\right)} \cong 1. \quad (15)$$

In this hierarchical parameterization, the mass matrix $M'_q$ in Eqn. (9) assumes the following form

$$M'_q = \begin{pmatrix} m_1 \xi_q & \sqrt{m_1 m_2 (1-\xi_q)(1+\zeta_q)}\, e^{i\alpha_q} & 0 \\ \sqrt{m_1 m_2 (1-\xi_q)(1+\zeta_q)}\, e^{-i\alpha_q} & \dfrac{\zeta_q m_3}{1+\zeta_q} & \dfrac{\sigma_q m_3 \sqrt{(\zeta_q + m_2/m_3)}}{1+\zeta_q}\, e^{i\beta_q} \\ 0 & \dfrac{\sigma_q m_3 \sqrt{(\zeta_q + m_2/m_3)}}{1+\zeta_q}\, e^{-i\beta_q} & \dfrac{m_3}{1+\zeta_q} \end{pmatrix}. \quad (16)$$

## 4. Obtaining the CKM matrix

One can now compute the quark mixing matrix or the CKM matrix through

$$V = V_{CKM} = O_u^\dagger P_u P_d^\dagger O_d. \quad (17)$$

In general, the various elements of the quark mixing matrix can be expressed

$$V_{iv} = O_{1i}^u O_{1v}^d e^{-i\varphi_1} + O_{2i}^u O_{2v}^d + O_{3i}^u O_{3v}^d e^{i\varphi_2}, \quad (18)$$

where the phases $\varphi_1 = \alpha_u - \alpha_d$ and $\varphi_2 = \beta_u - \beta_d$ are free parameters.

**4.1 CASE I:** Consider the weak basis representation $(M'_u, M'_d)$ given in Eqn. (6) such that

$$M'_u = \begin{pmatrix} e_u & |a_u|e^{i\alpha_u} & 0 \\ |a_u|e^{-i\alpha_u} & d_u & |b_u|e^{i\beta_u} \\ 0 & |b_u|e^{-i\beta_u} & c_u \end{pmatrix}, \quad M'_d = \begin{pmatrix} 0 & |a_d|e^{i\alpha_d} & 0 \\ |a_d|e^{-i\alpha_d} & d_d & |b_d|e^{i\beta_d} \\ 0 & |b_d|e^{-i\beta_d} & c_d \end{pmatrix}. \quad (19)$$

It is easy to check that for natural quark mass matrices, the parameter $\sigma_q \approx 1$. The various CKM matrix elements may then be expressed in terms of the quark mass ratios, $\xi_u, \zeta_u, \zeta_d$ and the phases $\phi_1$ and $\phi_2$, e.g.,

$$V_{ud} = e^{-i\phi_1} + \sqrt{\frac{m_u m_d}{m_c m_s}} \sqrt{\frac{(1-\xi_u)}{(1+\zeta_u)(1+\zeta_d)}}, \quad (20)$$

$$V_{us} = \sqrt{\frac{m_d}{m_s}} e^{-i\phi_1} - \sqrt{\frac{m_u}{m_c}} \sqrt{\frac{(1-\xi_u)}{(1+\zeta_u)(1+\zeta_d)}}, \quad (21)$$

$$V_{ub} = \sqrt{\frac{m_d m_s}{m_b^2}} \sqrt{\zeta_d} e^{-i\phi_1} + \sqrt{\frac{m_u}{m_c}} \sqrt{\frac{(1-\xi_u)}{(1+\zeta_u)(1+\zeta_d)}} \left[ \sqrt{\zeta_d} - \sqrt{\zeta_u} e^{i\phi_2} \right], \quad (22)$$

$$V_{cd} = \sqrt{\frac{m_u}{m_c}} \sqrt{(1-\xi_u)} e^{-i\phi_1} - \sqrt{\frac{m_d}{m_s}} \sqrt{\frac{1}{(1+\zeta_u)(1+\zeta_d)}} \left[ 1 + \sqrt{\zeta_u \zeta_d} e^{i\phi_2} \right], \quad (23)$$

$$V_{cs} = \sqrt{\frac{1}{(1+\zeta_u)(1+\zeta_d)}} \left[ 1 + \sqrt{\zeta_u \zeta_d} e^{i\phi_2} \right], \quad (24)$$

$$V_{cb} = -\sqrt{\frac{1}{(1+\zeta_u)(1+\zeta_d)}} \left[ \sqrt{\zeta_d} - \sqrt{\zeta_u} e^{i\phi_2} \right], \quad (25)$$

$$V_{td} = \sqrt{\frac{m_d}{m_s}} \sqrt{\frac{1}{(1+\zeta_u)(1+\zeta_d)}} \left[ \sqrt{\zeta_u} - \sqrt{\zeta_d} e^{i\phi_2} \right], \quad (26)$$

$$V_{ts} = -\sqrt{\frac{1}{(1+\zeta_u)(1+\zeta_d)}} \left[ \sqrt{\zeta_u} - \sqrt{\zeta_d} e^{i\phi_2} \right], \quad (27)$$

$$V_{tb} = \sqrt{\frac{1}{(1+\zeta_u)(1+\zeta_d)}} \left[ \sqrt{\zeta_u \zeta_d} + e^{i\phi_2} \right], \quad (28)$$

More explicitly, the four vital quark mixing parameters can be expressed as

$$|V_{us}| = \left| \sqrt{\frac{m_d}{m_s}} e^{-i\phi_1} - \sqrt{\frac{m_u}{m_c}} \sqrt{\frac{(1-\xi_u)}{(1+\zeta_u)(1+\zeta_d)}} \right|, \quad (29)$$

$$\frac{|V_{ub}|}{|V_{cb}|} = \left| \sqrt{\frac{m_u}{m_c}} \sqrt{(1-\xi_u)} + \sqrt{\frac{m_d m_s}{m_b^2}} \sqrt{(1+\zeta_u)(1+\zeta_d)} \frac{e^{-i\phi_1}}{(1-\sigma_d \sqrt{\zeta_u/\zeta_d}\, e^{i\phi_2})} \right|, \qquad (30)$$

$$|V_{cb}| = \left| \sqrt{\frac{1}{(1+\zeta_u)(1+\zeta_d)}} \left[ \sqrt{\zeta_d} - \sqrt{\zeta_u}\, e^{i\phi_2} \right] \right|, \qquad (31)$$

$$\beta = \beta_1 + \beta_2 = \arg\left( \sigma_d^2 - \sqrt{\frac{m_u m_s}{m_c m_d}} \sqrt{(1-\xi_u)}\, e^{-i(\phi_1+\phi_2)} \right) + \arg\left( \frac{\sqrt{\zeta_u/\zeta_d} - \sigma_u \sigma_d e^{i\phi_2}}{\sqrt{\zeta_u/\zeta_d} - (\sigma_u/\sigma_d) e^{i\phi_2}} \right) \qquad (32)$$

The above expressions are found to hold good within an error of less than a percent. It is interesting to observe the explicit contributions of the hierarchy characterizing parameters to the various mixing elements as well as the unitarity angle β. Furthermore, for Fritzsch-like texture 4 zero hermitian quark mass matrices, wherein $e_q = 0$ or $\xi_q = 0$, the expressions become much simpler. Explicitly, one observes that the parameter $\xi_u = e_u/m_u$ has no effect on $V_{cb}$. This is easy to interpret as the element $\xi_u$ does not invoke mixing among the second and third generation of quarks and hence does not appear in the expression for $V_{cb}$ given in Eqn. (31). Furthermore, from the Eqn. (29) it is also observed that $\xi_u$ does not make any significant contribution in the calculation of $V_{us}$ since $\sqrt{(m_u/m_c(1+\zeta_u)(1+\zeta_d))} \ll \sqrt{m_d/m_s}$. It is most likely that the non zero values of the parameters $\xi_q$ have implications, if any, for the smallest quark mixing angle $s_{13} \sim V_{ub}$ [11] or the CP asymmetry parameter $\sin 2\beta$.

**4.2 CASE II:** Consider the weak basis representation $(M'_u, M'_d)$ given in Eqn. (7) such that

$$M'_u = \begin{pmatrix} 0 & |a_u|e^{i\alpha_u} & 0 \\ |a_u|e^{-i\alpha_u} & d_u & |b_u|e^{i\beta_u} \\ 0 & |b_u|e^{-i\beta_u} & c_u \end{pmatrix}, \qquad M'_d = \begin{pmatrix} e_d & |a_d|e^{i\alpha_d} & 0 \\ |a_d|e^{-i\alpha_d} & d_d & |b_d|e^{i\beta_d} \\ 0 & |b_d|e^{-i\beta_d} & c_d \end{pmatrix}. \qquad (33)$$

In this case, the various CKM matrix elements can be expressed in terms of the quark mass ratios, $\xi_d, \zeta_u, \zeta_d$ and the phases $\phi_1$ and $\phi_2$, e.g.,

$$V_{ud} = e^{-i\phi_1} + \sqrt{\frac{m_u m_d}{m_c m_s}} \sqrt{\frac{(1-\xi_d)}{(1+\zeta_u)(1+\zeta_d)}}, \qquad (34)$$

$$V_{us} = \sqrt{\frac{m_d}{m_s}} \sqrt{(1-\xi_d)}\, e^{-i\phi_1} - \sqrt{\frac{m_u}{m_c}} \sqrt{\frac{1}{(1+\zeta_u)(1+\zeta_d)}}, \qquad (35)$$

$$V_{ub} = \sqrt{\frac{m_d m_s}{m_b^2}} \sqrt{\zeta_d} \sqrt{(1-\xi_d)}\, e^{-i\phi_1} + \sqrt{\frac{m_u}{m_c}} \sqrt{\frac{1}{(1+\zeta_u)(1+\zeta_d)}} \left[ \sqrt{\zeta_d} - \sqrt{\zeta_u}\, e^{i\phi_2} \right], \qquad (36)$$

$$V_{cd} = \sqrt{\frac{m_u}{m_c}} e^{-i\phi_1} - \sqrt{\frac{m_d}{m_s}} \sqrt{\frac{(1-\xi_d)}{(1+\zeta_u)(1+\zeta_d)}} \left[ 1 + \sqrt{\zeta_u \zeta_d} e^{i\phi_2} \right], \tag{37}$$

$$V_{cs} = \sqrt{\frac{1}{(1+\zeta_u)(1+\zeta_d)}} \left[ 1 + \sqrt{\zeta_u \zeta_d} e^{i\phi_2} \right], \tag{38}$$

$$V_{cb} = -\sqrt{\frac{1}{(1+\zeta_u)(1+\zeta_d)}} \left[ \sqrt{\zeta_d} - \sqrt{\zeta_u} e^{i\phi_2} \right], \tag{39}$$

$$V_{td} = \sqrt{\frac{m_d}{m_s}} \sqrt{\frac{(1-\xi_d)}{(1+\zeta_u)(1+\zeta_d)}} \left[ \sqrt{\zeta_u} - \sqrt{\zeta_d} e^{i\phi_2} \right], \tag{40}$$

$$V_{ts} = -\sqrt{\frac{1}{(1+\zeta_u)(1+\zeta_d)}} \left[ \sqrt{\zeta_u} - \sqrt{\zeta_d} e^{i\phi_2} \right], \tag{41}$$

$$V_{tb} = \sqrt{\frac{1}{(1+\zeta_u)(1+\zeta_d)}} \left[ \sqrt{\zeta_u \zeta_d} + e^{i\phi_2} \right]. \tag{42}$$

And the four vital quark mixing parameters can be expressed as

$$|V_{us}| = \left| \sqrt{\frac{m_d}{m_s}} \sqrt{(1-\xi_d)} e^{-i\phi_1} - \sqrt{\frac{m_u}{m_c}} \sqrt{\frac{1}{(1+\zeta_u)(1+\zeta_d)}} \right|, \tag{43}$$

$$\frac{|V_{ub}|}{|V_{cb}|} = \left| \sqrt{\frac{m_u}{m_c}} + \sqrt{\frac{m_d m_s}{m_b^2}} \sqrt{(1-\xi_d)(1+\zeta_u)(1+\zeta_d)} \frac{e^{-i\phi_1}}{(1-\sigma_d \sqrt{\zeta_u/\zeta_d}\, e^{i\phi_2})} \right|, \tag{44}$$

$$|V_{cb}| = \left| \sqrt{\frac{1}{(1+\zeta_u)(1+\zeta_d)}} \left[ \sqrt{\zeta_d} - \sqrt{\zeta_u} e^{i\phi_2} \right] \right|, \tag{45}$$

$$\beta = \beta_1 + \beta_2 = \arg\left( \sigma_d^2 - \sqrt{\frac{m_u m_s}{m_c m_d}} \sqrt{\frac{1}{(1-\xi_d)}} e^{-i(\phi_1+\phi_2)} \right) + \arg\left( \frac{\sqrt{\zeta_u/\zeta_d} - \sigma_u \sigma_d e^{i\phi_2}}{\sqrt{\zeta_u/\zeta_d} - (\sigma_u/\sigma_d) e^{i\phi_2}} \right), \tag{46}$$

It may be emphasized that using the parameterization of Eqn. (14), the explicit contributions of the hierarchy characterizing parameters to the various CKM mixing elements as well as the unitarity angle β can be precisely illustrated. Here again it is noted that the parameter $\xi_d = e_d/m_d$ does not appear in the expression for $V_{cb}$ given in Eqn. (45) and hence its non-zero values do not contribute in the evaluation of $V_{cb}$. Furthermore, the occurrence of element $\xi_d$ in the expression for $V_{us}$ in Eqn. (43), is not expected to make significant contributions to it, since the phase $\phi_1$ is a free parameter and can accommodate the effects of non-zero $\xi_d$. However, it should be interesting to investigate the implications of this additional diagonal element $\xi_d$, if any, for $V_{ub}$ and sin 2β.

## 5. Inputs and Constraints used for the Analysis

In order to investigate the implications of the elements $\xi_q$ for the quark mixing and CP violation parameters, the following ranges of quark masses [10] at the energy scale of $M_z$, have been adopted e.g.,

$$m_u = 0.8 - 1.8 \text{ MeV}, \quad m_d = 1.7 - 4.2 \text{ MeV}, \quad m_s = 40.0 - 71.0 \text{ MeV},$$
$$m_c = 0.6 - 0.7 \text{ GeV}, \quad m_b = 2.8 - 3.0 \text{ GeV}, \quad m_t = 169.5 - 175.5 \text{ GeV}. \tag{47}$$

The values of the vital quark mixing parameters viz. $V_{us}$, $V_{cb}$, $V_{ub}$ have been constrained using the Eqn. (2) along with $\sin 2\beta = 0.679 \pm 0.020$ or $\beta = 20.6° - 22.17°$ [11].

## 6. Results:

**6.1 Case I:** To start with, we first investigate the parameter space allowed to the additional element $\xi_u$ by the above inputs and constraints. In this regard, figure 1 shows that $\xi_u$ is permitted to span approximately the entire phenomenological range $(0 < \xi_u < 0.98)$ allowed to it by Eqn. (11). Hence all values of $e_u$ between $0 < e_u < 0.98\, m_u$ are allowed by experimental results. It is observed that the non-zero values of $\xi_u$ do not appear to have any significant implications for $V_{us}$, $V_{cb}$, $V_{ub}$ and CP asymmetry parameter $\sin 2\beta$ as the complete allowed ranges for these parameters in Eqn. (2) can be reproduced with the parameter space available to $e_u$. As a consequence of this it is observed that non-zero $\xi_u$ and hence $e_u$ do not make any significant contributions in reproducing the experimental results.

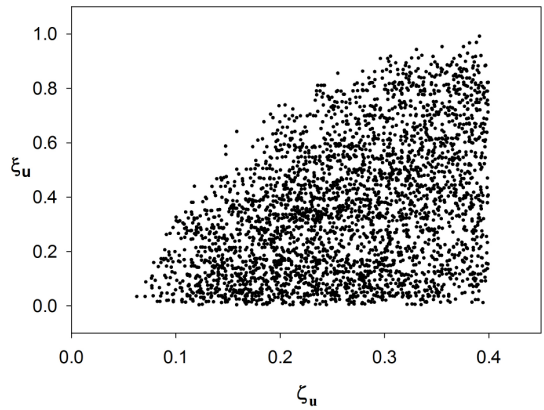

Figure 1: Plot showing the allowed parameter spaces for $\xi_u$ and $\zeta_u$ in case I.

**6.2 Case II:** Figure 2 depicts the values of the parameter $\xi_d$ allowed by the inputs and constraints. It is observed that, unlike the previous case, $\xi_d$ is not permitted to span the complete phenomenological range viz. $(0 < \xi_d < 1)$ but is constrained to have values between $0 < \xi_d < 0.43$ only. This can be understood from Eqn. (43) due to the occurrence of $\sqrt{(1-\xi_d)}$ in the first term which is the dominant term involved in the determination of $V_{us}$ and hence values of $\xi_d > 0.43$ are forbidden by the experimental values of $V_{us}$ due to limitations in the allowed values of the ratio $\sqrt{m_d/m_s} = 0.2253 - 0.2351$ [14]. However, the values of $\xi_d < 0.43$ can be explained due to the freedom of the parameter space available to the free parameter $\phi_1$ in accommodating the experimental values of $V_{us}$. This is shown in the figure 3 wherein it is observed that as $\xi_d$ increases beyond zero, the viable range of free parameter $\phi_1$ also increases from 79°-91° to 156°-178°.

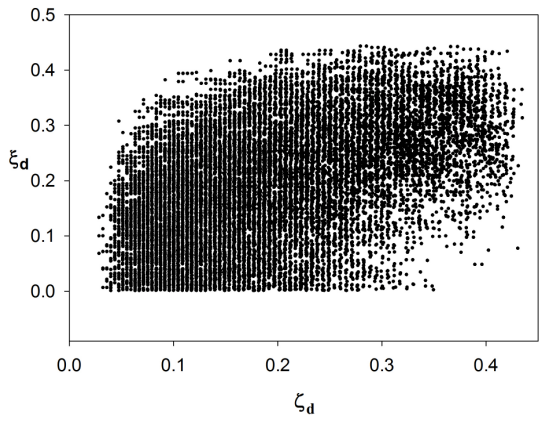

Figure 2: Plot showing the allowed parameter spaces for $\xi_d$ and $\zeta_d$ in the case II.

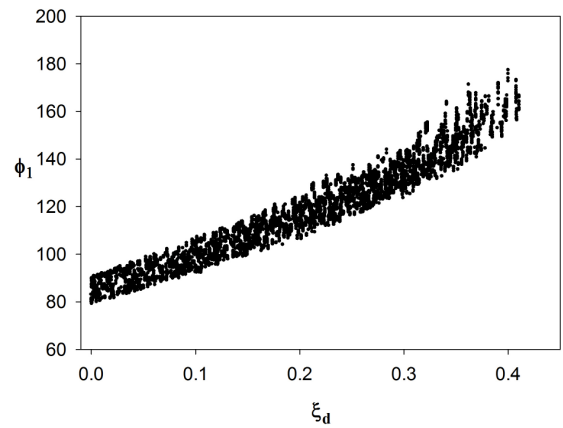

Figure 3: Plot showing the variation of the phase $\phi_1$ with increase in $\xi_d$ in the case II.

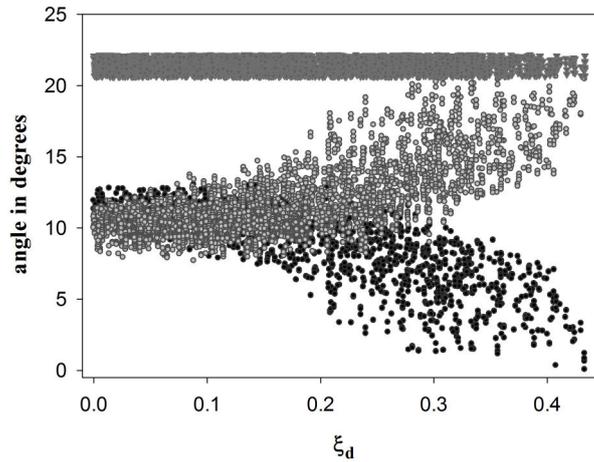

Figure 4: Plot showing the variation of $\beta_1$ and $\beta_2$ of Eqn, (46) with increase in $\xi_d$ in the case II. The black dots represent $\beta_1$, the gray dots represent $\beta_2$ and the gray triangles between 20° - 25° represent total CP asymmetry angle β.

The relative contributions of the two terms $\beta_1$ and $\beta_2$ contributing to the CP asymmetry angle β in Eqn. (46) as a function of $\xi_d$ are expressed in figure 4. It is interesting to observe that as $\xi_d$ increases the contribution of $\beta_1$ decreases (black dots) from around 12° to 0° whereas that of $\beta_2$ increases (grey dots) from around 10° to 22°, so as to reproduce the complete experimental range of the CP asymmetry angle β = 20.6° - 22.17° [11] (gray triangles). Similar observations are noted for the corresponding values of $\beta_1$ and $\beta_2$ as a function of increasing $\xi_u$ in the case I. In this case also, it is observed that $\xi_d$ and hence $e_d$ do not make any significant contributions in reproducing the complete range of the experimental values of $V_{us}$, $V_{cb}$ and more importantly $V_{ub}$ and CP asymmetry parameter sin 2β. Furthermore, it is also observed from the figures 1 and 2 that the smallest allowed values for $\zeta_u$ and $\zeta_d$ for the WB texture 3 zero mass matrices are 0.06 and 0.04 respectively, in agreement with the corresponding ones obtained for the texture 4 zero Fritzsch-like hermitian quark mass matrices [4], [9].

## 7. Conclusions

It is noted that the WB texture 3 zero hermitian quark mass matrices in Eqns. (6) and (7) differ from Fritzsch-like hermitian texture 4 zero quark mass matrices in Eqn. (8) due to the presence of non-zero $e_q = (M_q)_{11}$ elements in these. It has been observed that the condition of hermicity on WB texture 3 zero quark mass matrices requires that these additional parameters $e_q$ can have very small values only, viz. $-m_2 < e_q < m_1$ implying that the WB texture 3 zero quark mass matrices also exhibit natural structure given in Eqn. (4). Introducing the hierarchical parameterization, using $\xi_q$ and $\zeta_q$, we have been able to illustrate the explicit relationship among the observed hierarchies in the quark mass matrices and the CKM matrix through precise mathematical relations wherein all the CKM elements as well as the unitarity angle β are completely expressible in terms of the quark masses, $\xi_q$, $\zeta_q$, and the phases $\phi_1$ and $\phi_2$. Furthermore, it is interestingly observed that one can reproduce the entire 1σ range of all the vital quark mixing parameters viz. $V_{us}$, $V_{cb}$, $V_{ub}$ and sin 2β for the complete allowed range of $e_q$ or $\xi_q = e_q/m_1$. It may be noted that we have not included the Renormalization Group Equation (RGE) dependence of the considered quark masses in Eqn. (47) on the mod values of CKM matrix elements given in Eqn. (2) from Ref. [11]. It is trivial to check that a 10% - 20% variation arising from the calculation of the light quark masses at $M_Z$ using RGE [10] leaves the results unchanged in view of the large parameter space available to other parameters viz. $\zeta_u$, $\zeta_d$, $\phi_1$ and $\phi_2$ [6], [8] defined through Eqn. (11) and depicted through figure 3 in section 6.2. As a result, the broad conclusions drawn in the current manuscript are expected to be largely unaffected by the inclusion of these. This suggests that the additional parameters $e_q$ in the WB texture 3 zero quark mass matrices may be considered to be redundant as these do not appear to make any significant contributions in reproducing the quark mixing data as well as observed CP violation implying that the phenomenological difference between the 3-zero (weak basis choice) and 4-zero (one assumption added) textures is actually small, and that the latter is very close to reality. One can thus conclude that for the purpose of accommodating the quark mixing data, without loss of generality, the Fritzsch-like texture 4 zero hermitian mass matrices can be considered to be equivalent to the WB texture 3 zero hermitian quark mass matrices, since the $e_q = (M_q)_{11}$ elements in these do not appear to have any physical implications for the quark mixing data and observed CP violation.

**Acknowledgements**
The author would like to thank the Director, Rayat Institute of Engineering and Information Technology for providing the facilities to work.